\documentclass[%
reprint,
superscriptaddress,
showkeys,
showpacs,preprintnumbers,
 amsmath,amssymb,
floatfix,
]{revtex4-1}

\usepackage[USenglish]{babel} 
\usepackage[T1]{fontenc}
\usepackage[ansinew]{inputenc}
\usepackage{tabularx}
\usepackage{lmodern} 
\usepackage{graphicx} %%For loading graphic files
\usepackage{amsmath}
\usepackage{amsthm}
\usepackage{amsfonts}
\usepackage{gensymb}
\usepackage{booktabs}
\usepackage{filecontents}
\usepackage{color}
\usepackage{mhchem}
\usepackage{upgreek}

\usepackage{mathtools}
\usepackage{booktabs} % Tarjoaa \toprule-, \midrule- ja \bottomrule-komennot taulukkoja varten
\usepackage{siunitx} 
\begin{document}

%\preprint{AIP/123-QED}

\title{Superconducting tunnel junction fabrication on three-dimensional topography via direct laser writing}
% Force line breaks with \\

\author{Samuli Heiskanen}
\author{Ilari J. Maasilta}%
 \email{maasilta@jyu.fi}
\affiliation{Nanoscience Center, Department of Physics, University of Jyvaskyla, P. O. Box 35, FIN-40014 Jyv\"askyl\"a, Finland 
%\\This line break forced with \textbackslash\textbackslash
}%

% \homepage{http://www.Second.institution.edu/~Charlie.Author.}
%\affiliation{%
%Second institution and/or address%\\This line break forced% with \\
%}%

\date{\today}% It is always \today, today,

\begin{abstract}
Superconducting junctions are widely used in multitude of applications ranging from quantum information science and sensing to solid-state cooling. Traditionally, such devices must be fabricated on flat substrates using standard lithographic techniques. In this study, we demonstrate a highly versatile method that allows for superconducting junctions to be fabricated on a more complex topography. It is based on maskless direct laser writing (DLW) two-photon lithography, which allows writing in 3D space. We show that high-quality normal metal-insulator-superconductor (NIS) tunnel junctions can be fabricated on top of a 20 $\mu$m tall three-dimensional topography. 
Combined with more advanced resist coating methods, this technique could allow sub-micron device fabrication on almost any type of topography in the future.
\end{abstract}

\maketitle

%\keywords{Superconducting tunnel junction, NIS junction, direct laser writing, two-photon absorption, patterning on large topographical features, positive-tone resist, lift-off}

%\section{Introduction}

Superconducting tunnel junctions have many applications for example in sensing \cite{SQUID}, quantum information \cite{WilhelmClarke,Wendin_2017} and nanoscale thermal sciences \cite{Giazotto_heikkila}. Their fabrication is done using the well-established techniques of ultraviolet photolithography or electron-beam lithography, which work fine except for one major flaw: the patterning is always done on flat 2D substrates. This is a limitation for more advanced device designs, which may in some cases require or benefit from the junctions sitting on some elevated platforms or at the bottom of a trench, for example. It is the purpose of this study to demonstrate a technique that facilitates such more advanced, yet high quality superconducting junction fabrication on substrates of varying topography.        

As the first proof-of-principle demonstration, in this study we fabricate and characterize normal metal-insulator-superconductor (NIS) tunnel junctions \cite{giaever} on complex topography. This choice is based on two reasons: i) The electrical response of a NIS junction is a sensitive probe of the quality of the superconducting material and the insulating barrier, and ii) our own immediate application is in sensitive local thermometry of complex nanoscale 3D structures. The fabrication technique is, however, perfectly suited for the fabrication of SIS or SNS Josephson junctions, as well.

%Normal metal-insulator-superconductor (NIS) tunnel junctions were studied first by Fisher and Giaever, who used Al$/$\ce{Al2O3}$/$Pb devices with Pb being the superconductor and Al the normal metal \cite{fishergiaever}. Nowadays NIS junction devices are used for many applications usually as pairs, which are called SINIS junctions. 
NIS junctions are particularly suited for low-temperature thermometry, because the current through an NIS junction has a strong temperature dependence at energies close to the superconducting gap \cite{Giazotto_heikkila,Koppinen2009,PhysRevB.85.014519}, and they can even be considered in some cases as primary electron thermometers \cite{PhysRevApplied.4.034001}. Due to their sensitivity and the possibility to fabricate them in sub-micron scales, they are  excellent local sensors for heat transfer measurements \cite{Pekolanature,jenni,Zen_ncomms,Tian_2019}, can be operated fast in microsecond time scales with a microwave readout \cite{schmidt,PhysRevApplied.3.014007}, and could work as the sensor element in bolometric radiation detection \cite{martinis,nahum_Xray,heikkila,kuzmin2019} or in direct measurement of temperature fluctuations \cite{pekola2020}. Typically, the superconducting material used is Al, limiting the use of NIS devices to below 1 K. However, by using higher transition temperature superconductors, the temperature range of NIS thermometers has been extended with Nb \cite{minna,Juhani}, NbN \cite{nbn}, TaN \cite{TaN} and TiN \cite{andrii}.  

 Other possible applications of NIS junctions are in metrology \cite{pekolaturnstile,RevModPhys.85.1421}, thermal rectification \cite{Giazotto} and electronic cooling \cite{martiniscool,muhonen}, with demonstrations of cooling of macroscopic and mesoscopic platforms \cite{ZhangUllom,Nguyen,Prunnila}, nanoscale beams \cite{PhysRevLett.102.165502,Juha}, radiation detectors \cite{Ullom}, remote devices via photons \cite{PhysRevLett.102.200801, Mottonen} and quantum information circuit components \cite{Qbit_sinis}. %It is expected that many of these applications could benefit from the capability to integrate NIS devices into more complex   

%The most common fabrication method for NIS junction devices is electron beam lithography, due to it's very high resolution. This allows fabrication of small area junctions with high tunneling resistance. With electron beam lithography the use of co-polymers is standardized, which allows easy fabrication of undercuts and bridge structures for shadow masking. UV lithography is also a viable method but then you lose the higher resolution and maskless nature of electron beam lithography. Also undercut structures are more difficult to fabricate. However, both these methods are limited to 2D structures.

In this letter, we demonstrate high quality microscale Cu-AlO$_x$-Al  NIS junction fabrication on three-dimensional (3D) topography, using direct laser writing (DLW), which is a recently developed fabrication technique based on two-photon absorption (TPA), originally developed for writing arbitrary 3D polymer structures from negative photoresists \cite{sun,kawata,deubel}. In contrast, here we use DLW to develop a positive photoresist without a photomask in combination with metal deposition and lift-off, as introduced in Refs. \cite{acssensors.6b00469,samuli}.       %The DLW two-photon lithography method used for the fabrication was developed from our earlier work \cite{samuli}. 
%We demonstrate fast fabrication of microscale NIS junctions (writing speeds $\sim xxx \mu$m/s), %on any substrate, but due to the inherent 3D capabilities of two-photon lithography instruments, it also allows
 %submicron fabrication 
Junctions were fabricated both on flat substrates and, in particular, on top of a 20 micrometer tall 3D platform. Such devices are impossible to make by any other, more standard lithographic techniques.  Electrical characterization of the junctions demonstrate that the junction quality is high, as the standard junction theory fits the data extremely well, with a low level of excess sub-gap current. The temperature sensitivity extends to the lowest refrigerator temperatures used, demonstrating the application potential for thermometry and cooling of complex 3D device platforms.       %Similar methods have been used for photonic metamaterials \cite{Gansel1513}, magnetic microrobots \cite{smll.201302856} and magnetic nanostructures \cite{Williams2018}  by combining positive-tone resists with electrodeposition of metals and lift-off. In addition, recently a study \cite{acssensors.6b00469} pointed out the relevance of two-photon absorption (TPA) direct writing in fabrication of purely 2D, large area metallic nanostructures.

%Another way to use DLW for fabrication on 3D structures is writing metallic structures directly using photoreduction \cite{waller2018,hirt2017}. However, this method is not suitable for advanced devices like tunnel junctions, because high-quality pure materials with low roughness are required and with this method the quality of the produced films is limited by the chemistry involved. This is not an issue for our new method since it relies on established photoresists and evaporation techniques. The photoreduction method also has a limited range of materials (Ag, Au, and Cu), which makes it impossible to produce superconducting devices for example.

%\section{Fabrication methods}
%\label{fabrication}

\begin{figure}
    \centering
    \includegraphics[width=\columnwidth]{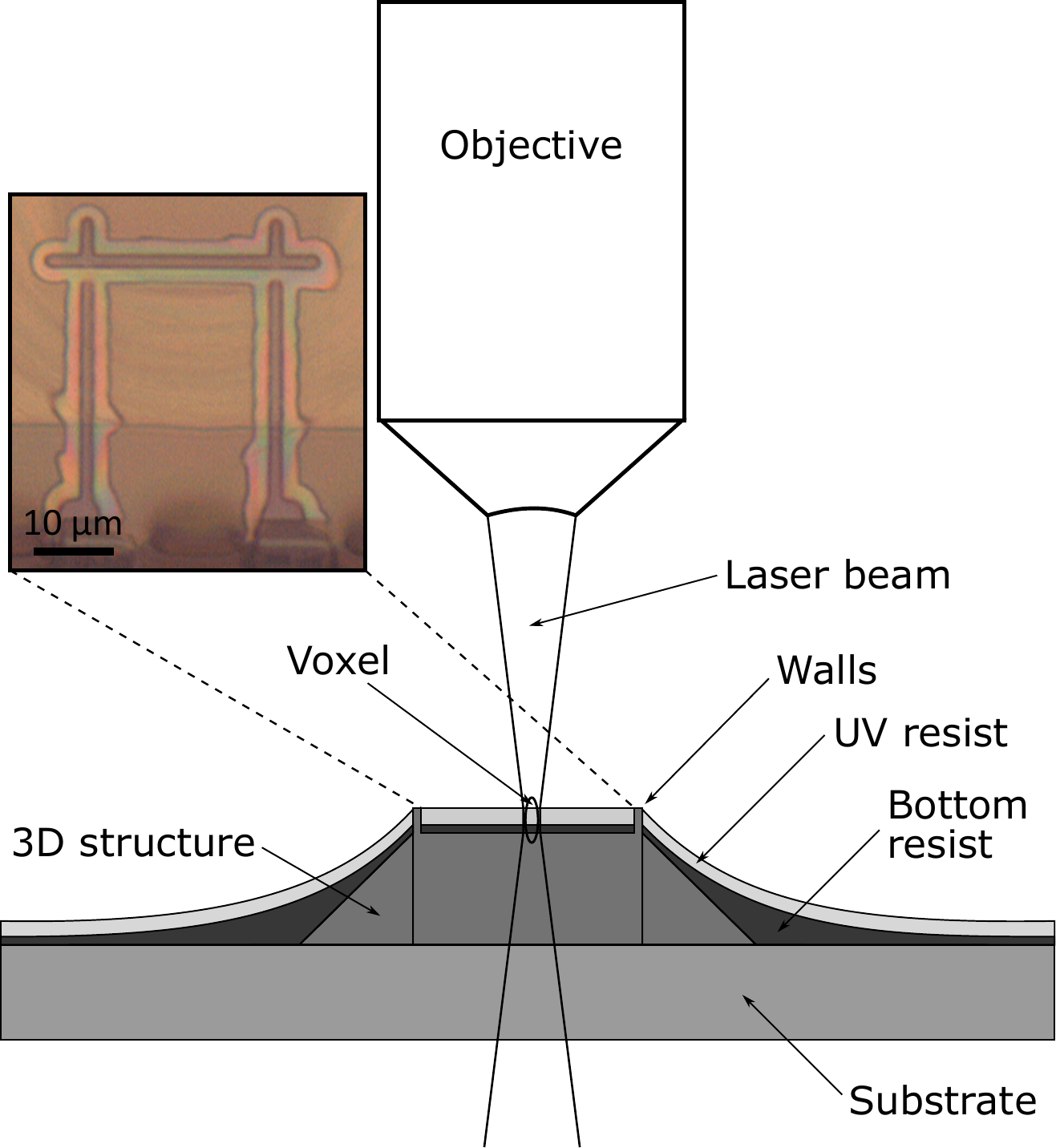}
    \caption{Schematic of the two-photon laser exposure method used for patterning the junctions. An air-gap objective is used in order to expose from the top side of the 3D structure without being in contact with the substrate. The focus spot of the laser beam travels in 3D space, following the existing topography on the front surface, where the TPA process forms the voxel. Inset: Optical micrograph of the exposed pattern on the 3D structure after development. The line width here is around 2.5 $\mu$m giving a junction area of 6.25 $\mu\text{m}^2$. A large $\sim 2$ $\mu$m  undercut and a bridge structure is visible.  
		%The ex of the undercut can be controlled with  development time}
		}
    \label{fig:spin}
\end{figure}

%Sample fabrication started with creating the 3D topography for the subsequent junction fabrication. 
The complex topography used is a 3D cuboid structure with area 100 $\mu$m x 100 $\mu$m  and height 20 $\mu$m with sloped ramps (Fig. ~\ref{fig:spin}), fabricated from a negative photoresist (IP-Dip, Nanoscribe GmbH) on nitridized Si substrate using DLW (Nanoscribe Photonic Professional), similarly to the cuboid structure in Ref. \cite{samuli}.  %system. This system uses a pulsed (80 MHz) laser with a 780 nm wavelength for 3D lithography via TPA. %A three axis piezo-electrical stage is used for accurate motion of the substrate in three dimensions. The cuboid structure, with dimensions 100 $\mu$m x 100 $\mu$m x 20 $\mu$m, is fabricated on an SiN substrate using the dip-in method, where an objective (NA = 1.3, 100x) is dipped into a liquid negative-tone photoresist (IP-Dip, Nanoscribe GmbH) for the exposure and photopolymerization. After the sample is exposed it is developed in PGMEA (propylene glycol monomethyl ether acetate), which removes the excess liquid resist revealing the drawn structure.
An additional 200 nm AlOx capping layer was evaporated on the whole cuboid structure to strengthen it and to obtain more homogeneous and flat surfaces. % properties on the sample. %The evaporation was done with an electron beam vacuum evaporation system at a pressure of $2*10^{-5}$ mbar. 

\begin{figure}
    \centering
    \includegraphics[width=0.7\columnwidth]{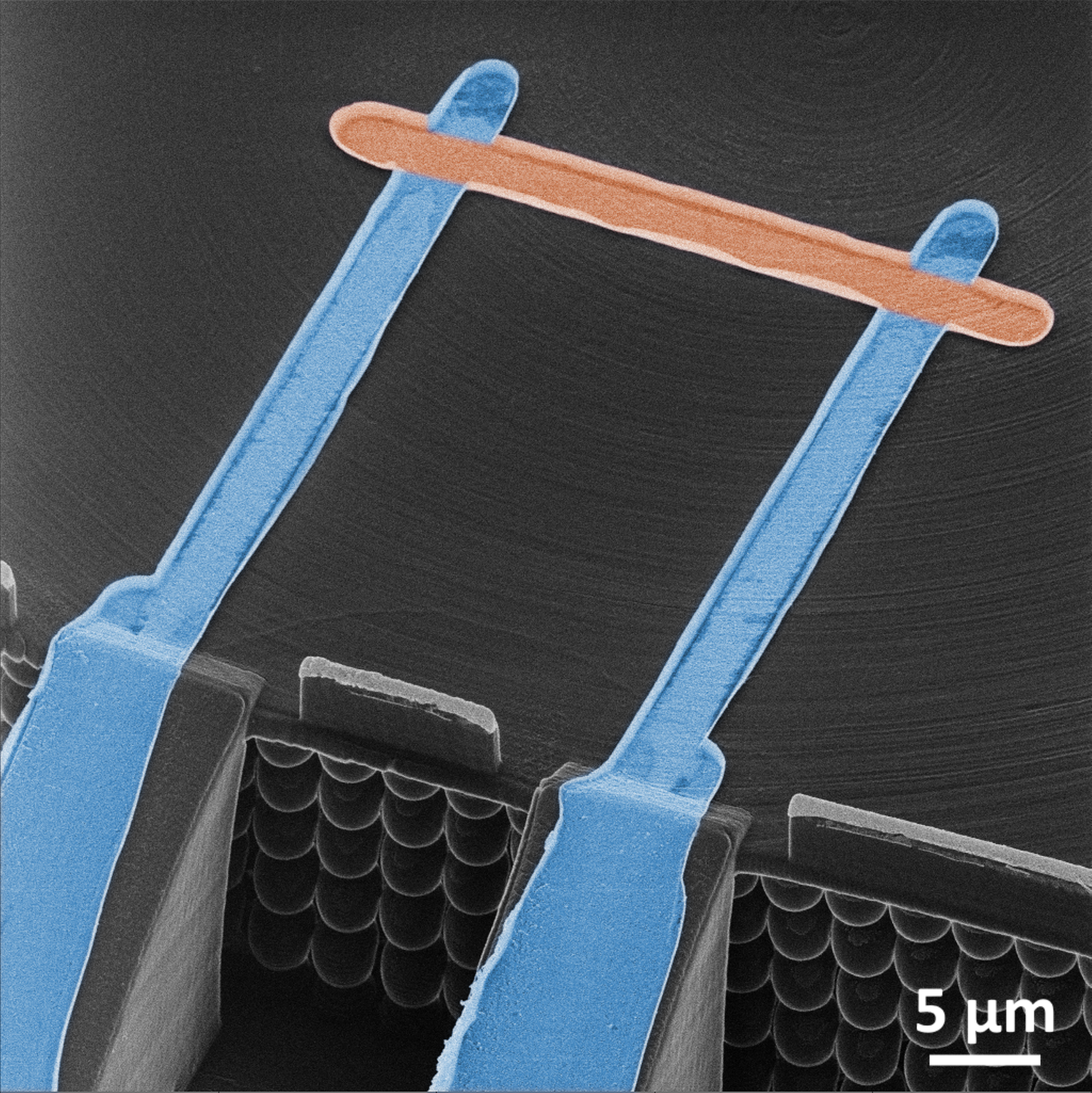}
    \caption{False color helium ion micrograph of a finished SINIS junction on the 3D structure (blue=Al, red=Cu). For this sample, the line width is around 3 $\mu$m, giving a junction area of 9 $\mu\text{m}^2$. %Clearly there is also metal on top of the small walls that go around the structure, but it is not connected to the junction circuit, due to the design of the structure.}
		}
    \label{fig:finished}
\end{figure}

The actual junction fabrication proceeds as follows: the sample is spin coated first with a bottom resist (AR-BR 5460, Allresist GmbH), which was diluted with propylene glycol methyl ether acetate (PGMEA) to a solids content of 9 \% to get a 450 nm thick film on a flat surface. This layer will allow for an undercut in the resist structure. We do not bake the bottom resist, but instead, keep it in vacuum overnight to remove the solvent. This allows for deeper undercuts and makes it possible to fabricate bridge structures for the subsequent shadow evaporation step. Then, the sample is spin coated with the UV  sensitive positive-tone resist (AR-P 3540, Allresist GmbH), to a nominal thickness of 2.8 $\mu$m  and baked at 100 $^{\circ}$  C for 140 s. To allow for the resists to stay on top of the cuboid structure during spinning, small vertical walls of height 3.5 $\mu$m were incorporated at the edges of the 3D platform, as shown in Figures~\ref{fig:spin} and ~\ref{fig:finished}. %The height of these walls is about 3.5 $\mu$m, %and they are required for a successful spin coating, 
Such walls could be left out if direct spray coating were used \cite{spray}. %However, this method was not available to us, so spin coating was used. 
%For later samples 

%\begin{figure}
 %   \centering
  %  \includegraphics[width=\columnwidth]{developed2.png}
  %  \caption{Helium ion microscope image of the exposed pattern on the 3D structure after development. The line width here is around 2.5 $\mu$m giving a junction area of 6,25 $\mu\text{m}^2$. The depth of the undercut can be controlled with development time. Here the development time was short giving almost no undercut.} 
 %   \label{fig:developed}
%\end{figure}

The lithographic exposure of the junction geometry is done with the same Nanoscribe Photonic Professional tool (pulse rate 80 MHz, wavelength 780 nm) that was used for the fabrication of the topography.  Because now solid resists are used (as opposed to liquid), the objective (63x, NA = 0.75) cannot be in contact with the resist, and an air gap to the sample remains, as %Air-Gap method, where the laser beam comes from the side of the topography, 
shown in Figure~\ref{fig:spin}. %This was actually the only possible choice, because the topography requires the laser to be focused from above the sample and the objective cannot be in contact with the sample. For this method we used a 63x objective with an NA of 0.75. 
A writing speed $v = 25 \mu$m/s and laser power $P =$ 4 mW was used for the exposure on top of the platform, producing a single-pass line width $\approx 1 \mu$m through the the whole top resist layer. The size of the voxel (line resolution) can easily be controlled with the power and speed, scaling as $P^2/v$ \cite{samuli}, with possible speeds up to a few mm/s. As the aspect ratio of the voxel is naturally high (height:width 5:1), it is straightforward to expose microns thick resist layers on one pass. Moreover, a large voxel height allows for significant, even micron scale variations of the actual resist thickness, which invariably exist when spin coating such tall topographical features. Even thicker resists can be exposed by layering the writing pattern, and this technique was used in areas near the base of the platform structure where the resist was thickest.  

In our setup, the focus of the laser cannot automatically follow the topography and thus the height changes need to be included in the pattern of the laser writing path. This means that the dimensions of the topography need to be known in advance for this method. Fortunately, this information does not need to be very precise due to the size of the voxel. The alignment of the sample was done with the integrated optical microscope of the Photonic Professional tool.

After the exposure, the sample is developed in a 1:1 mixture of AR 300-47 developer (Allresist GmbH) and deionized water, and rinsed in deionized water. With the two layers of resists used, the depth of the undercut can be controlled simply with the development time. This is because the bottom resist is not exposed at all and just slowly dissolves into the developer. The development times used for our samples were around 15 seconds, producing a micron-scale undercut profile (inset, Figure~\ref{fig:spin}). %For the structures with deeper undercut the development time was reduced to 15 seconds due to the faster dissolving rate of the non baked bottom resist. 

After the patterning and development, the metals are evaporated using an ultra-high vacuum electron-beam evaporator. First, a 30 nm layer of aluminum is evaporated  
 along the patterned leads that climb the ramps. This was done from two opposing directions, in four different steps from an angle increasing from 60 to 70 degrees. This sequence ensured better coverage over the roughness of the underlying structure. The aluminum is then thermally oxidized in 200 mbar of pure oxygen for 9 minutes, producing an AlOx tunnel barrier. %layer is created by keeping the sample  
Finally, the sample is rotated 90 $^{\circ}$  and a 60 nm copper layer is evaporated similarly to the aluminum. Angle evaporation is also used for the copper, so that no copper will land  on the aluminum leads on top of the platform. After the evaporation, lift-off is done with hot acetone. A helium ion micrograph of a finished double junction SINIS device is shown in Figure~\ref{fig:finished}.

\begin{figure*}
\centering
\includegraphics[width =\textwidth]{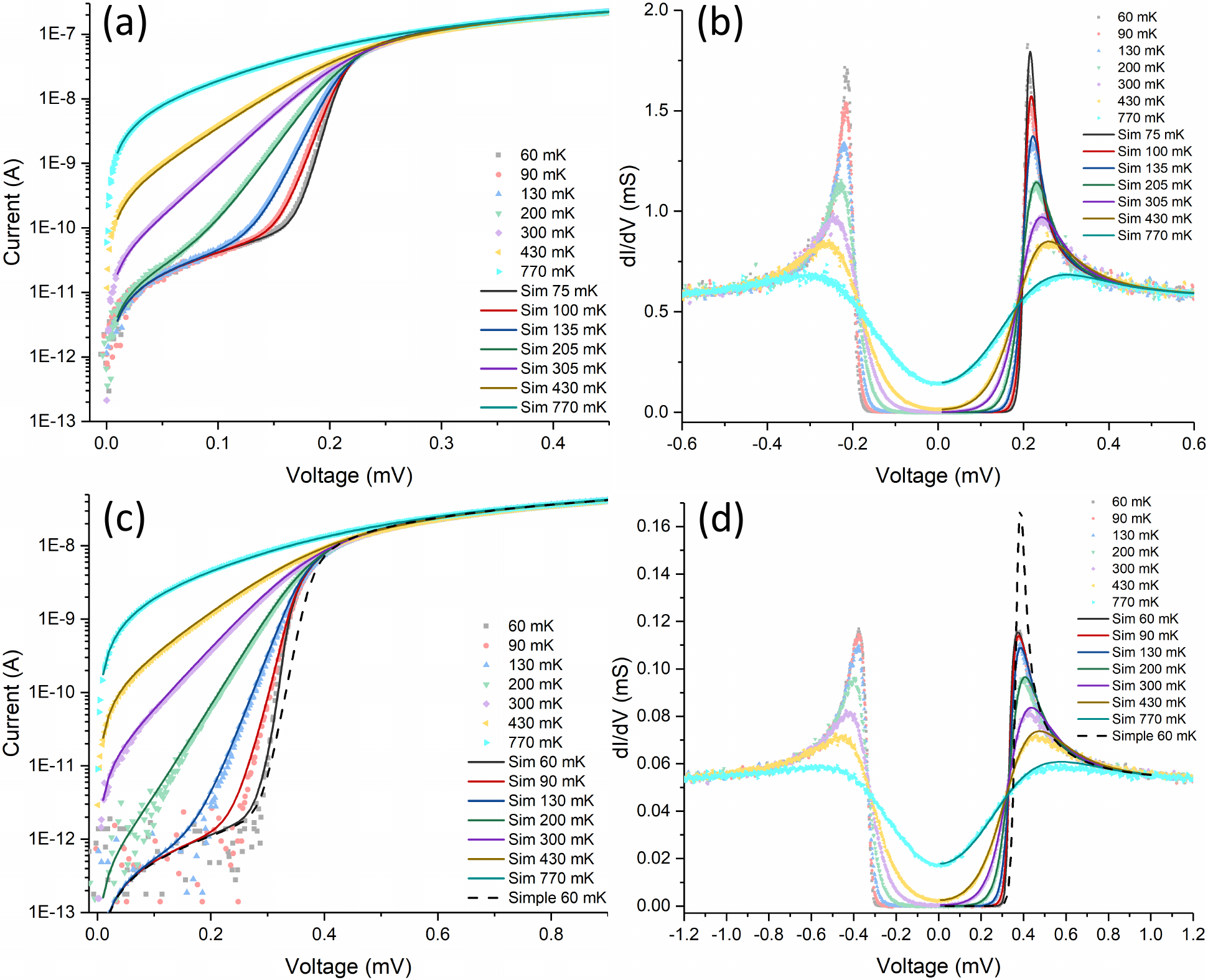}
\caption{The measured low-temperature (a) $I$-$V$  and (b) $dI/dV$-$V$ characteristics for the DLW fabricated NIS junctions on a flat substrate (symbols) (junction area 5 $\mu\text{m}^2$, $R_T \sim 2 $k$\Omega$). Theory fits to Eq. (1) and its voltage derivative are shown as lines, with the electron temperature as a fitting parameter. The experimental $dI/dV$ data was produced by numerical differentiation of the $I-V$ curve. %(a) IV curve measured from a single NIS junction of a junction area of 5 $\mu\text{m}^2$ on a flat SiN substrate. %The junction consisted of aluminum, $\ce{Al2O3}$ and copper, and it had a junction area of 5 $\mu\text{m}^2$. 
%(b) Calculated differential curve for the NIS junction. 
(c) and (d): The same for a SINIS junction pair on top of the fabricated topography (junction area 4 $\mu\text{m}^2$, $R_{T,\rm{tot}} \sim 20 $k$\Omega$).  The solid lines are the theoretical results based on Eq. (1) with the thermal modeling described in the text. The dashed line shows the simple theory result of Eq. (1) with fixed $T$ (no thermal model) for the data at bath temperature 60 mK. }
\label{fig:graphs}
\end{figure*}

%\section{Measurements}
%\label{measurements}

%According to the BCS theory, 
The current through a high-quality NIS junction is given by the expression
\begin{equation}
	\label{eq:isinis}
	I =\frac{1}{2eR_T}\int_{-\infty}^{\infty}d\epsilon N_S(\epsilon)[f_N(\epsilon-eV)-f_N(\epsilon+eV)],
\end{equation}
\noindent
where $R_T$ is the tunneling resistance of the junction, $N_S(\epsilon)$ is the density of states of the superconductor and $f_N$ is the Fermi-Dirac distribution in the normal metal \cite{Giazotto_heikkila,Koppinen2009}. For the density of states, we use the expression taking into account non-idealities
\begin{equation}
	\label{eq:dynes}
	N_S(\epsilon, T_S)=\left|\text{Re}\left(\frac{\epsilon+i\Gamma}{\sqrt{(\epsilon+i\Gamma)^2-\Delta^2(T_S)}}\right)\right|,
\end{equation}
\noindent
where $\Gamma$ is the so called Dynes parameter and $\Delta(T_S)$ the temperature dependent superconducting energy gap. $\Gamma$ describes in general the non-ideal broadening of the density of states \cite{dynes} due to barrier and material non-idealities. For the case of high-quality Al junctions, it was ultimately shown to result from environmentally photon-assisted tunneling events \cite{Pekola_dynes}. 
% This parameter is set-up dependent in the case of photon-assisted tunneling \cite{Giazotto_heikkila}. The $\Delta(T_S)$ here, is the superconducting energy gap with temperature dependence. This form for the density of states is particularly appropriate for Al junctions, in which it 

To study the properties of NIS junctions fabricated with the new DLW method, we first fabricated junctions on flat nitridized Si substrates and measured them using a $^{3}$He/$^{4}$He dilution refrigerator with a base temperature of 60 mK. %did measurements for simpler samples. For these samples we used the same fabrication method to fabricate NIS junctions on a flat SiN substrate. 
Examples of sets of $I$-$V$ and $dI/dV$-$V$  measurements on such a single NIS junction device with $R_T = 2$ k$\Omega$, as a function of the bath temperature, are shown in Figure~\ref{fig:graphs}(a)-(b). They are plotted together with theory fits based on Eqs. (1) and (2), with the electron temperature fitted but constant for each curve, and $\Delta$ and $R_T$ fitted but kept constant for the whole set.   We can see that the fits are extremely good for both the $I$-$V$ and the $dI/dV$ data, demonstrating that the DLW method can produce junctions of equal quality to standard lithography.    %with measurement results for one NIS junction with simulated curves for every other temperature . The simulations were done with the BCS theory using Equation (\ref{eq:isinis}) at constant temperatures and the simulations fit the measured data well. 

The fitted temperatures match almost exactly the measured bath temperatures, except at the lowest temperatures, where the fitted electron temperatures are slightly higher. This is a well-known effect, caused by absorbed spurious thermal radiation from the higher temperature stages of the cryostat, in combination with the strong thermal decoupling of the electron system from the lattice at the lowest temperatures \cite{Wellstood,jenni,PhysRevLett.102.165502}. From the fits we determined the zero temperature superconducting gap of the aluminum to be $\Delta(0)=$ 0.208 meV, agreeing with previous results on a thin film Al of similar thickness deposited in the same evaporator \cite{PhysRevLett.102.165502}. The Dynes parameter for this device was $\Gamma/\Delta(0) = 6.5 \times 10^{-4}$, a value roughly consistent with what has been measured in the same setup before \cite{PhysRevLett.102.165502}, and low enough to allow for efficient NIS cooling \cite{TaN,PhysRevLett.92.056804}. However, for this particular device, electronic cooling was not observed due to the large size of the normal metal electrode in the single junction geometry, and due to the lack of normal metal quasiparticle traps contacting the Al electrode \cite{traps}.

%Then we did measurements for a junction on top of the fabricated topography. The measurements were otherwise exactly the same as for the simpler sample, but now we were measuring a SINIS junction pair. 
In Figure~\ref{fig:graphs}(c) and (d), similar measurements are shown for a double junction SINIS device fabricated on top of the 3D topography (Fig. \ref{fig:finished}). This time, the fits with the simplest constant temperature model do not produce good results anymore. However, by incorporating the thermal resistance of the small normal metal island due to the electron-phonon interaction and the effect of Joule self-heating \cite{Giazotto_heikkila,PhysRevB.85.014519}, the fits become nearly perfect. The impact of the self-heating is particularly noticeable as the rounding of the differential conductance peaks of the lowest bath temperature data [Figure~\ref{fig:graphs}(d)]. %The This heating is probably due to the fact that the junction is well isolated from the thermal bath on top of the fabricated 3D structure. Due to the heating the fits could not be made with the same simple model as for the results in Figure~\ref{fig:graphs}a-b. Now we had to use a thermal model which takes the heating into account by balancing the heat flows out of and into the normal metal island.

%The heat current $\dot{Q}_i$ through each junction $(i=L,R)$ is given by the following expressions \cite{PhysRevB.85.014519}:

%\begin{equation} \label{eq:heat_l}
%\begin{split}
%	\dot{Q}_L(V_L,T_N,T_S)=&\frac{1}{e^2R_L}\int_{-\infty}^{\infty}d\epsilon(\epsilon+eV_L)N_S(\epsilon,T_S) \\
%	&*[f_S(\epsilon,T_S)-f_N(\epsilon+eV_L,T_N)],
%\end{split}
%\end{equation}

%\begin{equation} \label{eq:heat_r}
%\begin{split}
%	\dot{Q}_R(V_R,T_N,T_S)=&\frac{1}{e^2R_R}\int_{-\infty}^{\infty}d\epsilon(\epsilon)N_S(\epsilon+eV_R,T_S) \\
%	&*[f_N(\epsilon,T_N)-f_S(\epsilon+eV_R,T_S)].
%\end{split}
%\end{equation}

%\noindent
%Knowing the heat currents, the total heat escaping the normal metal can be calculated from

%\begin{equation}
%	\label{eq:escape}
%	P_{es}=-\dot{Q}_L+\dot{Q}_R.
%\end{equation}

%We can also write an expression for the inflow of heat from the surroundings, which is

%\begin{equation}
%	\label{eq:inflow}
%	P_{in}=B(T_{bath}^n-T_N^n)+\beta[P_T+I(V_L+V_R)]+I^2R_N,
%\end{equation}

%\noindent

In the thermal model, described in more detail in Ref. \cite{PhysRevB.85.014519}, we used the usual relation \cite{Wellstood} $P = \Sigma V(T_e^5-T_p^5)$ for the power flow between electrons and phonons for thick metal films on bulky substrates, where $V$ is the normal metal volume, and a typical value $\Sigma = 2 \times 10^9$ W/(K$^5$m$^3$) for the electron-phonon coupling strength in Cu was used \cite{Giazotto_heikkila}. In addition, the modeling included the parameter $\beta = 0.25$, giving the fraction of dissipated power that back-flows from the superconducting electrodes into the normal metal \cite{fisherullom}. This value is higher than what is typically observed ($\beta < 0.1$) for NIS junctions that are optimized for cooling \cite{muhonen,PhysRevLett.102.165502}, which is expected, as we have a thin superconducting film without quasiparticle trapping.  %The first term gives the direct heating coming from the substrate ($T_{bath}$) and the exponent $n$ in the term depends on the disorder level and phonon dimensionality of the sample \cite{PhysRevLett.99.145503}, but is usually $n=5$. The second term describes the heat flowing back from the superconducting electrodes and the third term gives the Joule heating of the normal metal.

%In dynamic equilibrium the escaping heat is balanced by the inflow of heat, which means that by balancing Equations (\ref{eq:escape}) and (\ref{eq:inflow}) we can find the temperature of the normal metal $T_N$. For the simulations in Figure~\ref{fig:graphs}c-d we used the parameters $B=5*10^{-8}$, $\beta=1$ and $n=5$. 
For this SINIS junction, the Dynes parameter has a quite a low value $\Gamma/\Delta(0)=9 \times 10^{-5}$, which is much lower than for the single NIS junction sample. This improvement is most likely due to the higher value of $R_T$, leading to less efficient absorption of environmental radiation power and thus smaller current due to photon-assisted tunneling. The superconducting gap is now $\Delta(0) =0.181$ meV, smaller than for the simpler NIS junction but still in agreement with literature \cite{RevModPhys.35.1}.

%\section{Conclusions}

In this study, the most challenging step of the process was in fact the resist coating. With the simple spin coating used here, we had to to fabricate ramps for the wiring and additional walls to contain the resist on top of the platform topography. However, we anticipate that by using more advanced resist coating methods, such as spray coating, the fabrication could be done over almost any type of topography, with much steeper vertical gradients. 

In conclusion, we have shown that direct laser writing based on two-photon lithography can be used for the fabrication of advanced, high-quality superconducting tunnel junction devices, with our case study concentrating on normal metal-insulator-superconductor (NIS) tunnel junctions. This method can be used for fast maskless fabrication over large areas down to sub-micron scales, and it is very versatile since it uses established positive photoresists and the well-known angle evaporation and lift-off techniques. Even more importantly, we have demonstrated that the method can be used for superconducting tunnel junction fabrication on highly varying 20 $\mu$m tall 3D topography, something that is impossible with standard lithography. %In particular, in our case study here was for normal metal-insulator-superconductor (NIS) tunnel junctions,   % More importantly we have proven that this method can be used for fabrication of advanced devices over large topographies. 
%The NIS tunnel junction measurements show normal junction behavior, and the superconducting gap of the aluminum agrees with literature, for junctions both on the SiN substrate and on the fabricated 20 $\mu$m high topography. This tells us that the process can produce pure high-quality metal films even on the topography, so fabrication of all kinds of devices on large topographies should be possible. 
Such fabrication opens up a multitude of possibilities to integrate superconducting devices with 3D geometries for advanced applications in ultrasensitive radiation detectors and quantum information processing, for example.

\begin{acknowledgments}
This study was supported by the Academy of Finland project number 298667. We thank Mr. Zhuoran Geng for enlightening discussions and technical assistance. 
\end{acknowledgments}

%\newpage
%\bibliography{reference}

\begin{thebibliography}{51}%
\makeatletter
\providecommand \@ifxundefined [1]{%
 \@ifx{#1\undefined}
}%
\providecommand \@ifnum [1]{%
 \ifnum #1\expandafter \@firstoftwo
 \else \expandafter \@secondoftwo
 \fi
}%
\providecommand \@ifx [1]{%
 \ifx #1\expandafter \@firstoftwo
 \else \expandafter \@secondoftwo
 \fi
}%
\providecommand \natexlab [1]{#1}%
\providecommand \enquote  [1]{``#1''}%
\providecommand \bibnamefont  [1]{#1}%
\providecommand \bibfnamefont [1]{#1}%
\providecommand \citenamefont [1]{#1}%
\providecommand \href@noop [0]{\@secondoftwo}%
\providecommand \href [0]{\begingroup \@sanitize@url \@href}%
\providecommand \@href[1]{\@@startlink{#1}\@@href}%
\providecommand \@@href[1]{\endgroup#1\@@endlink}%
\providecommand \@sanitize@url [0]{\catcode `\\12\catcode `\$12\catcode
  `\&12\catcode `\#12\catcode `\^12\catcode `\_12\catcode `\%12\relax}%
\providecommand \@@startlink[1]{}%
\providecommand \@@endlink[0]{}%
\providecommand \url  [0]{\begingroup\@sanitize@url \@url }%
\providecommand \@url [1]{\endgroup\@href {#1}{\urlprefix }}%
\providecommand \urlprefix  [0]{URL }%
\providecommand \Eprint [0]{\href }%
\providecommand \doibase [0]{http://dx.doi.org/}%
\providecommand \selectlanguage [0]{\@gobble}%
\providecommand \bibinfo  [0]{\@secondoftwo}%
\providecommand \bibfield  [0]{\@secondoftwo}%
\providecommand \translation [1]{[#1]}%
\providecommand \BibitemOpen [0]{}%
\providecommand \bibitemStop [0]{}%
\providecommand \bibitemNoStop [0]{.\EOS\space}%
\providecommand \EOS [0]{\spacefactor3000\relax}%
\providecommand \BibitemShut  [1]{\csname bibitem#1\endcsname}%
\let\auto@bib@innerbib\@empty
%</preamble>
\bibitem [{\citenamefont {Clarke}\ and\ \citenamefont
  {Braginski}(2004)}]{SQUID}%
  \BibitemOpen
  \bibinfo {editor} {\bibfnamefont {J.}~\bibnamefont {Clarke}}\ and\ \bibinfo
  {editor} {\bibfnamefont {A.~I.}\ \bibnamefont {Braginski}},\ eds.,\
  \href@noop {} {\emph {\bibinfo {title} {The SQUID handbook}}}\ (\bibinfo
  {publisher} {Wiley-VCH},\ \bibinfo {address} {Weinheim},\ \bibinfo {year}
  {2004})\BibitemShut {NoStop}%
\bibitem [{\citenamefont {Clarke}\ and\ \citenamefont
  {Wilhelm}(2008)}]{WilhelmClarke}%
  \BibitemOpen
  \bibfield  {author} {\bibinfo {author} {\bibfnamefont {J.}~\bibnamefont
  {Clarke}}\ and\ \bibinfo {author} {\bibfnamefont {F.~K.}\ \bibnamefont
  {Wilhelm}},\ }\href@noop {} {\bibfield  {journal} {\bibinfo  {journal}
  {Nature}\ }\textbf {\bibinfo {volume} {453}},\ \bibinfo {pages} {1031}
  (\bibinfo {year} {2008})}\BibitemShut {NoStop}%
\bibitem [{\citenamefont {Wendin}(2017)}]{Wendin_2017}%
  \BibitemOpen
  \bibfield  {author} {\bibinfo {author} {\bibfnamefont {G.}~\bibnamefont
  {Wendin}},\ }\href {\doibase 10.1088/1361-6633/aa7e1a} {\bibfield  {journal}
  {\bibinfo  {journal} {Reports on Progress in Physics}\ }\textbf {\bibinfo
  {volume} {80}},\ \bibinfo {pages} {106001} (\bibinfo {year}
  {2017})}\BibitemShut {NoStop}%
\bibitem [{\citenamefont {Giazotto}\ \emph {et~al.}(2006)\citenamefont
  {Giazotto}, \citenamefont {Heikkil\"a}, \citenamefont {Luukanen},
  \citenamefont {Savin},\ and\ \citenamefont {Pekola}}]{Giazotto_heikkila}%
  \BibitemOpen
  \bibfield  {author} {\bibinfo {author} {\bibfnamefont {F.}~\bibnamefont
  {Giazotto}}, \bibinfo {author} {\bibfnamefont {T.~T.}\ \bibnamefont
  {Heikkil\"a}}, \bibinfo {author} {\bibfnamefont {A.}~\bibnamefont
  {Luukanen}}, \bibinfo {author} {\bibfnamefont {A.~M.}\ \bibnamefont {Savin}},
  \ and\ \bibinfo {author} {\bibfnamefont {J.~P.}\ \bibnamefont {Pekola}},\
  }\href {\doibase 10.1103/RevModPhys.78.217} {\bibfield  {journal} {\bibinfo
  {journal} {Rev. Mod. Phys.}\ }\textbf {\bibinfo {volume} {78}},\ \bibinfo
  {pages} {217} (\bibinfo {year} {2006})}\BibitemShut {NoStop}%
\bibitem [{\citenamefont {Giaever}(1960)}]{giaever}%
  \BibitemOpen
  \bibfield  {author} {\bibinfo {author} {\bibfnamefont {I.}~\bibnamefont
  {Giaever}},\ }\href {\doibase 10.1103/PhysRevLett.5.147} {\bibfield
  {journal} {\bibinfo  {journal} {Phys. Rev. Lett.}\ }\textbf {\bibinfo
  {volume} {5}},\ \bibinfo {pages} {147} (\bibinfo {year} {1960})}\BibitemShut
  {NoStop}%
\bibitem [{\citenamefont {Koppinen}\ \emph {et~al.}(2009)\citenamefont
  {Koppinen}, \citenamefont {K{\"u}hn},\ and\ \citenamefont
  {Maasilta}}]{Koppinen2009}%
  \BibitemOpen
  \bibfield  {author} {\bibinfo {author} {\bibfnamefont {P.~J.}\ \bibnamefont
  {Koppinen}}, \bibinfo {author} {\bibfnamefont {T.}~\bibnamefont {K{\"u}hn}},
  \ and\ \bibinfo {author} {\bibfnamefont {I.~J.}\ \bibnamefont {Maasilta}},\
  }\href {\doibase 10.1007/s10909-009-9861-7} {\bibfield  {journal} {\bibinfo
  {journal} {Journal of Low Temperature Physics}\ }\textbf {\bibinfo {volume}
  {154}},\ \bibinfo {pages} {179} (\bibinfo {year} {2009})}\BibitemShut
  {NoStop}%
\bibitem [{\citenamefont {Chaudhuri}\ and\ \citenamefont
  {Maasilta}(2012)}]{PhysRevB.85.014519}%
  \BibitemOpen
  \bibfield  {author} {\bibinfo {author} {\bibfnamefont {S.}~\bibnamefont
  {Chaudhuri}}\ and\ \bibinfo {author} {\bibfnamefont {I.~J.}\ \bibnamefont
  {Maasilta}},\ }\href {\doibase 10.1103/PhysRevB.85.014519} {\bibfield
  {journal} {\bibinfo  {journal} {Phys. Rev. B}\ }\textbf {\bibinfo {volume}
  {85}},\ \bibinfo {pages} {014519} (\bibinfo {year} {2012})}\BibitemShut
  {NoStop}%
\bibitem [{\citenamefont {Feshchenko}\ \emph {et~al.}(2015)\citenamefont
  {Feshchenko}, \citenamefont {Casparis}, \citenamefont {Khaymovich},
  \citenamefont {Maradan}, \citenamefont {Saira}, \citenamefont {Palma},
  \citenamefont {Meschke}, \citenamefont {Pekola},\ and\ \citenamefont
  {Zumb\"uhl}}]{PhysRevApplied.4.034001}%
  \BibitemOpen
  \bibfield  {author} {\bibinfo {author} {\bibfnamefont {A.~V.}\ \bibnamefont
  {Feshchenko}}, \bibinfo {author} {\bibfnamefont {L.}~\bibnamefont
  {Casparis}}, \bibinfo {author} {\bibfnamefont {I.~M.}\ \bibnamefont
  {Khaymovich}}, \bibinfo {author} {\bibfnamefont {D.}~\bibnamefont {Maradan}},
  \bibinfo {author} {\bibfnamefont {O.-P.}\ \bibnamefont {Saira}}, \bibinfo
  {author} {\bibfnamefont {M.}~\bibnamefont {Palma}}, \bibinfo {author}
  {\bibfnamefont {M.}~\bibnamefont {Meschke}}, \bibinfo {author} {\bibfnamefont
  {J.~P.}\ \bibnamefont {Pekola}}, \ and\ \bibinfo {author} {\bibfnamefont
  {D.~M.}\ \bibnamefont {Zumb\"uhl}},\ }\href {\doibase
  10.1103/PhysRevApplied.4.034001} {\bibfield  {journal} {\bibinfo  {journal}
  {Phys. Rev. Applied}\ }\textbf {\bibinfo {volume} {4}},\ \bibinfo {pages}
  {034001} (\bibinfo {year} {2015})}\BibitemShut {NoStop}%
\bibitem [{\citenamefont {Meschke}\ \emph {et~al.}(2006)\citenamefont
  {Meschke}, \citenamefont {Guichard},\ and\ \citenamefont
  {Pekola}}]{Pekolanature}%
  \BibitemOpen
  \bibfield  {author} {\bibinfo {author} {\bibfnamefont {M.}~\bibnamefont
  {Meschke}}, \bibinfo {author} {\bibfnamefont {W.}~\bibnamefont {Guichard}}, \
  and\ \bibinfo {author} {\bibfnamefont {J.~P.}\ \bibnamefont {Pekola}},\
  }\href@noop {} {\bibfield  {journal} {\bibinfo  {journal} {Nature (London)}\
  }\textbf {\bibinfo {volume} {444}},\ \bibinfo {pages} {187} (\bibinfo {year}
  {2006})}\BibitemShut {NoStop}%
\bibitem [{\citenamefont {Karvonen}\ and\ \citenamefont
  {Maasilta}(2007)}]{jenni}%
  \BibitemOpen
  \bibfield  {author} {\bibinfo {author} {\bibfnamefont {J.~T.}\ \bibnamefont
  {Karvonen}}\ and\ \bibinfo {author} {\bibfnamefont {I.~J.}\ \bibnamefont
  {Maasilta}},\ }\href {\doibase 10.1103/PhysRevLett.99.145503} {\bibfield
  {journal} {\bibinfo  {journal} {Phys. Rev. Lett.}\ }\textbf {\bibinfo
  {volume} {99}},\ \bibinfo {pages} {145503} (\bibinfo {year}
  {2007})}\BibitemShut {NoStop}%
\bibitem [{\citenamefont {Zen}\ \emph {et~al.}(2014)\citenamefont {Zen},
  \citenamefont {Puurtinen}, \citenamefont {Isotalo}, \citenamefont
  {Chaudhuri},\ and\ \citenamefont {Maasilta}}]{Zen_ncomms}%
  \BibitemOpen
  \bibfield  {author} {\bibinfo {author} {\bibfnamefont {N.}~\bibnamefont
  {Zen}}, \bibinfo {author} {\bibfnamefont {T.~A.}\ \bibnamefont {Puurtinen}},
  \bibinfo {author} {\bibfnamefont {T.~J.}\ \bibnamefont {Isotalo}}, \bibinfo
  {author} {\bibfnamefont {S.}~\bibnamefont {Chaudhuri}}, \ and\ \bibinfo
  {author} {\bibfnamefont {I.~J.}\ \bibnamefont {Maasilta}},\ }\href@noop {}
  {\bibfield  {journal} {\bibinfo  {journal} {Nature Communications}\ }\textbf
  {\bibinfo {volume} {5}},\ \bibinfo {pages} {3435} (\bibinfo {year}
  {2014})}\BibitemShut {NoStop}%
\bibitem [{\citenamefont {Tian}\ \emph {et~al.}(2019)\citenamefont {Tian},
  \citenamefont {Puurtinen}, \citenamefont {Geng},\ and\ \citenamefont
  {Maasilta}}]{Tian_2019}%
  \BibitemOpen
  \bibfield  {author} {\bibinfo {author} {\bibfnamefont {Y.}~\bibnamefont
  {Tian}}, \bibinfo {author} {\bibfnamefont {T.~A.}\ \bibnamefont {Puurtinen}},
  \bibinfo {author} {\bibfnamefont {Z.}~\bibnamefont {Geng}}, \ and\ \bibinfo
  {author} {\bibfnamefont {I.~J.}\ \bibnamefont {Maasilta}},\ }\href {\doibase
  10.1103/PhysRevApplied.12.014008} {\bibfield  {journal} {\bibinfo  {journal}
  {Phys. Rev. Applied}\ }\textbf {\bibinfo {volume} {12}},\ \bibinfo {pages}
  {014008} (\bibinfo {year} {2019})}\BibitemShut {NoStop}%
\bibitem [{\citenamefont {Schmidt}\ \emph {et~al.}(2005)\citenamefont
  {Schmidt}, \citenamefont {Lehnert}, \citenamefont {Clark}, \citenamefont
  {Duncan}, \citenamefont {Irwin}, \citenamefont {Miller},\ and\ \citenamefont
  {Ullom}}]{schmidt}%
  \BibitemOpen
  \bibfield  {author} {\bibinfo {author} {\bibfnamefont {D.~R.}\ \bibnamefont
  {Schmidt}}, \bibinfo {author} {\bibfnamefont {K.~W.}\ \bibnamefont
  {Lehnert}}, \bibinfo {author} {\bibfnamefont {A.~M.}\ \bibnamefont {Clark}},
  \bibinfo {author} {\bibfnamefont {W.~D.}\ \bibnamefont {Duncan}}, \bibinfo
  {author} {\bibfnamefont {K.~D.}\ \bibnamefont {Irwin}}, \bibinfo {author}
  {\bibfnamefont {N.}~\bibnamefont {Miller}}, \ and\ \bibinfo {author}
  {\bibfnamefont {J.~N.}\ \bibnamefont {Ullom}},\ }\href
  {http://scitation.aip.org/content/aip/journal/apl/86/5/10.1063/1.1855411}
  {\bibfield  {journal} {\bibinfo  {journal} {Applied Physics Letters}\
  }\textbf {\bibinfo {volume} {86}},\ \bibinfo {eid} {053505} (\bibinfo {year}
  {2005})}\BibitemShut {NoStop}%
\bibitem [{\citenamefont {Gasparinetti}\ \emph {et~al.}(2015)\citenamefont
  {Gasparinetti}, \citenamefont {Viisanen}, \citenamefont {Saira},
  \citenamefont {Faivre}, \citenamefont {Arzeo}, \citenamefont {Meschke},\ and\
  \citenamefont {Pekola}}]{PhysRevApplied.3.014007}%
  \BibitemOpen
  \bibfield  {author} {\bibinfo {author} {\bibfnamefont {S.}~\bibnamefont
  {Gasparinetti}}, \bibinfo {author} {\bibfnamefont {K.~L.}\ \bibnamefont
  {Viisanen}}, \bibinfo {author} {\bibfnamefont {O.-P.}\ \bibnamefont {Saira}},
  \bibinfo {author} {\bibfnamefont {T.}~\bibnamefont {Faivre}}, \bibinfo
  {author} {\bibfnamefont {M.}~\bibnamefont {Arzeo}}, \bibinfo {author}
  {\bibfnamefont {M.}~\bibnamefont {Meschke}}, \ and\ \bibinfo {author}
  {\bibfnamefont {J.~P.}\ \bibnamefont {Pekola}},\ }\href {\doibase
  10.1103/PhysRevApplied.3.014007} {\bibfield  {journal} {\bibinfo  {journal}
  {Phys. Rev. Applied}\ }\textbf {\bibinfo {volume} {3}},\ \bibinfo {pages}
  {014007} (\bibinfo {year} {2015})}\BibitemShut {NoStop}%
\bibitem [{\citenamefont {Nahum}\ and\ \citenamefont
  {Martinis}(1993)}]{martinis}%
  \BibitemOpen
  \bibfield  {author} {\bibinfo {author} {\bibfnamefont {M.}~\bibnamefont
  {Nahum}}\ and\ \bibinfo {author} {\bibfnamefont {J.~M.}\ \bibnamefont
  {Martinis}},\ }\href {\doibase 10.1063/1.110237} {\bibfield  {journal}
  {\bibinfo  {journal} {Applied Physics Letters}\ }\textbf {\bibinfo {volume}
  {63}},\ \bibinfo {pages} {3075} (\bibinfo {year} {1993})}\BibitemShut
  {NoStop}%
\bibitem [{\citenamefont {Nahum}\ and\ \citenamefont
  {Martinis}(1995)}]{nahum_Xray}%
  \BibitemOpen
  \bibfield  {author} {\bibinfo {author} {\bibfnamefont {M.}~\bibnamefont
  {Nahum}}\ and\ \bibinfo {author} {\bibfnamefont {J.~M.}\ \bibnamefont
  {Martinis}},\ }\href {\doibase 10.1063/1.113723} {\bibfield  {journal}
  {\bibinfo  {journal} {Applied Physics Letters}\ }\textbf {\bibinfo {volume}
  {66}},\ \bibinfo {pages} {3203} (\bibinfo {year} {1995})}\BibitemShut
  {NoStop}%
\bibitem [{\citenamefont {Heikkil\"a}\ \emph {et~al.}(2018)\citenamefont
  {Heikkil\"a}, \citenamefont {Ojaj\"arvi}, \citenamefont {Maasilta},
  \citenamefont {Strambini}, \citenamefont {Giazotto},\ and\ \citenamefont
  {Bergeret}}]{heikkila}%
  \BibitemOpen
  \bibfield  {author} {\bibinfo {author} {\bibfnamefont {T.~T.}\ \bibnamefont
  {Heikkil\"a}}, \bibinfo {author} {\bibfnamefont {R.}~\bibnamefont
  {Ojaj\"arvi}}, \bibinfo {author} {\bibfnamefont {I.~J.}\ \bibnamefont
  {Maasilta}}, \bibinfo {author} {\bibfnamefont {E.}~\bibnamefont {Strambini}},
  \bibinfo {author} {\bibfnamefont {F.}~\bibnamefont {Giazotto}}, \ and\
  \bibinfo {author} {\bibfnamefont {F.~S.}\ \bibnamefont {Bergeret}},\ }\href
  {\doibase 10.1103/PhysRevApplied.10.034053} {\bibfield  {journal} {\bibinfo
  {journal} {Phys. Rev. Applied}\ }\textbf {\bibinfo {volume} {10}},\ \bibinfo
  {pages} {034053} (\bibinfo {year} {2018})}\BibitemShut {NoStop}%
\bibitem [{\citenamefont {Kuzmin}\ \emph {et~al.}(2019)\citenamefont {Kuzmin},
  \citenamefont {Pankratov}, \citenamefont {Gordeeva}, \citenamefont
  {Zbrozhek}, \citenamefont {Shamporov}, \citenamefont {.~Revin}, \citenamefont
  {Blagodatkin}, \citenamefont {Mas},\ and\ \citenamefont
  {de~Bernardis}}]{kuzmin2019}%
  \BibitemOpen
  \bibfield  {author} {\bibinfo {author} {\bibfnamefont {L.}~\bibnamefont
  {Kuzmin}}, \bibinfo {author} {\bibfnamefont {A.~L.}\ \bibnamefont
  {Pankratov}}, \bibinfo {author} {\bibfnamefont {A.~V.}\ \bibnamefont
  {Gordeeva}}, \bibinfo {author} {\bibfnamefont {V.~O.}\ \bibnamefont
  {Zbrozhek}}, \bibinfo {author} {\bibfnamefont {V.~A.}\ \bibnamefont
  {Shamporov}}, \bibinfo {author} {\bibfnamefont {L.~S.}\ \bibnamefont
  {.~Revin}}, \bibinfo {author} {\bibfnamefont {A.~V.}\ \bibnamefont
  {Blagodatkin}}, \bibinfo {author} {\bibfnamefont {S.}~\bibnamefont {Mas}}, \
  and\ \bibinfo {author} {\bibfnamefont {P.}~\bibnamefont {de~Bernardis}},\
  }\href@noop {} {\bibfield  {journal} {\bibinfo  {journal} {Communications
  Physics}\ }\textbf {\bibinfo {volume} {2}},\ \bibinfo {pages} {104} (\bibinfo
  {year} {2019})}\BibitemShut {NoStop}%
\bibitem [{\citenamefont {Karimi}\ \emph {et~al.}(2020)\citenamefont {Karimi},
  \citenamefont {Brange}, \citenamefont {Samuelsson},\ and\ \citenamefont
  {Pekola}}]{pekola2020}%
  \BibitemOpen
  \bibfield  {author} {\bibinfo {author} {\bibfnamefont {B.}~\bibnamefont
  {Karimi}}, \bibinfo {author} {\bibfnamefont {F.}~\bibnamefont {Brange}},
  \bibinfo {author} {\bibfnamefont {P.}~\bibnamefont {Samuelsson}}, \ and\
  \bibinfo {author} {\bibfnamefont {J.~P.}\ \bibnamefont {Pekola}},\
  }\href@noop {} {\bibfield  {journal} {\bibinfo  {journal} {Nat. Comm.}\
  }\textbf {\bibinfo {volume} {11}},\ \bibinfo {pages} {367} (\bibinfo {year}
  {2020})}\BibitemShut {NoStop}%
\bibitem [{\citenamefont {Nevala}\ \emph {et~al.}(2012)\citenamefont {Nevala},
  \citenamefont {Chaudhuri}, \citenamefont {Halkosaari}, \citenamefont
  {Karvonen},\ and\ \citenamefont {Maasilta}}]{minna}%
  \BibitemOpen
  \bibfield  {author} {\bibinfo {author} {\bibfnamefont {M.~R.}\ \bibnamefont
  {Nevala}}, \bibinfo {author} {\bibfnamefont {S.}~\bibnamefont {Chaudhuri}},
  \bibinfo {author} {\bibfnamefont {J.}~\bibnamefont {Halkosaari}}, \bibinfo
  {author} {\bibfnamefont {J.~T.}\ \bibnamefont {Karvonen}}, \ and\ \bibinfo
  {author} {\bibfnamefont {I.~J.}\ \bibnamefont {Maasilta}},\ }\href@noop {}
  {\bibfield  {journal} {\bibinfo  {journal} {Applied Physics Letters}\
  }\textbf {\bibinfo {volume} {101}},\ \bibinfo {eid} {112601} (\bibinfo {year}
  {2012})}\BibitemShut {NoStop}%
\bibitem [{\citenamefont {Julin}\ and\ \citenamefont
  {Maasilta}(2016)}]{Juhani}%
  \BibitemOpen
  \bibfield  {author} {\bibinfo {author} {\bibfnamefont {J.~K.}\ \bibnamefont
  {Julin}}\ and\ \bibinfo {author} {\bibfnamefont {I.~J.}\ \bibnamefont
  {Maasilta}},\ }\href {\doibase 10.1088/0953-2048/29/10/105003} {\bibfield
  {journal} {\bibinfo  {journal} {Superconductor Science and Technology}\
  }\textbf {\bibinfo {volume} {29}},\ \bibinfo {pages} {105003} (\bibinfo
  {year} {2016})}\BibitemShut {NoStop}%
\bibitem [{\citenamefont {Chaudhuri}\ \emph {et~al.}(2013)\citenamefont
  {Chaudhuri}, \citenamefont {Nevala},\ and\ \citenamefont {Maasilta}}]{nbn}%
  \BibitemOpen
  \bibfield  {author} {\bibinfo {author} {\bibfnamefont {S.}~\bibnamefont
  {Chaudhuri}}, \bibinfo {author} {\bibfnamefont {M.~R.}\ \bibnamefont
  {Nevala}}, \ and\ \bibinfo {author} {\bibfnamefont {I.~J.}\ \bibnamefont
  {Maasilta}},\ }\href@noop {} {\bibfield  {journal} {\bibinfo  {journal}
  {Applied Physics Letters}\ }\textbf {\bibinfo {volume} {102}},\ \bibinfo
  {eid} {132601} (\bibinfo {year} {2013})}\BibitemShut {NoStop}%
\bibitem [{\citenamefont {Chaudhuri}\ and\ \citenamefont
  {Maasilta}(2014)}]{TaN}%
  \BibitemOpen
  \bibfield  {author} {\bibinfo {author} {\bibfnamefont {S.}~\bibnamefont
  {Chaudhuri}}\ and\ \bibinfo {author} {\bibfnamefont {I.~J.}\ \bibnamefont
  {Maasilta}},\ }\href
  {http://scitation.aip.org/content/aip/journal/apl/104/12/10.1063/1.4869563}
  {\bibfield  {journal} {\bibinfo  {journal} {Applied Physics Letters}\
  }\textbf {\bibinfo {volume} {104}},\ \bibinfo {eid} {122601} (\bibinfo {year}
  {2014})}\BibitemShut {NoStop}%
\bibitem [{\citenamefont {Torgovkin}\ \emph {et~al.}(2015)\citenamefont
  {Torgovkin}, \citenamefont {Chaudhuri}, \citenamefont {Malm}, \citenamefont
  {Sajavaara},\ and\ \citenamefont {Maasilta}}]{andrii}%
  \BibitemOpen
  \bibfield  {author} {\bibinfo {author} {\bibfnamefont {A.}~\bibnamefont
  {Torgovkin}}, \bibinfo {author} {\bibfnamefont {S.}~\bibnamefont
  {Chaudhuri}}, \bibinfo {author} {\bibfnamefont {J.}~\bibnamefont {Malm}},
  \bibinfo {author} {\bibfnamefont {T.}~\bibnamefont {Sajavaara}}, \ and\
  \bibinfo {author} {\bibfnamefont {I.~J.}\ \bibnamefont {Maasilta}},\
  }\href@noop {} {\bibfield  {journal} {\bibinfo  {journal} {IEEE Trans. Appl.
  Supercond.}\ }\textbf {\bibinfo {volume} {25}},\ \bibinfo {pages} {1101604}
  (\bibinfo {year} {2015})}\BibitemShut {NoStop}%
\bibitem [{\citenamefont {Pekola}\ \emph {et~al.}(2008)\citenamefont {Pekola},
  \citenamefont {Vartiainen}, \citenamefont {M\"ott\"onen}, \citenamefont
  {Saira}, \citenamefont {Meschke},\ and\ \citenamefont
  {Averin}}]{pekolaturnstile}%
  \BibitemOpen
  \bibfield  {author} {\bibinfo {author} {\bibfnamefont {J.~P.}\ \bibnamefont
  {Pekola}}, \bibinfo {author} {\bibfnamefont {J.~J.}\ \bibnamefont
  {Vartiainen}}, \bibinfo {author} {\bibfnamefont {M.}~\bibnamefont
  {M\"ott\"onen}}, \bibinfo {author} {\bibfnamefont {O.-P.}\ \bibnamefont
  {Saira}}, \bibinfo {author} {\bibfnamefont {M.}~\bibnamefont {Meschke}}, \
  and\ \bibinfo {author} {\bibfnamefont {D.~V.}\ \bibnamefont {Averin}},\
  }\href@noop {} {\bibfield  {journal} {\bibinfo  {journal} {Nat. Phys.}\
  }\textbf {\bibinfo {volume} {4}},\ \bibinfo {pages} {120} (\bibinfo {year}
  {2008})}\BibitemShut {NoStop}%
\bibitem [{\citenamefont {Pekola}\ \emph {et~al.}(2013)\citenamefont {Pekola},
  \citenamefont {Saira}, \citenamefont {Maisi}, \citenamefont {Kemppinen},
  \citenamefont {M\"ott\"onen}, \citenamefont {Pashkin},\ and\ \citenamefont
  {Averin}}]{RevModPhys.85.1421}%
  \BibitemOpen
  \bibfield  {author} {\bibinfo {author} {\bibfnamefont {J.~P.}\ \bibnamefont
  {Pekola}}, \bibinfo {author} {\bibfnamefont {O.-P.}\ \bibnamefont {Saira}},
  \bibinfo {author} {\bibfnamefont {V.~F.}\ \bibnamefont {Maisi}}, \bibinfo
  {author} {\bibfnamefont {A.}~\bibnamefont {Kemppinen}}, \bibinfo {author}
  {\bibfnamefont {M.}~\bibnamefont {M\"ott\"onen}}, \bibinfo {author}
  {\bibfnamefont {Y.~A.}\ \bibnamefont {Pashkin}}, \ and\ \bibinfo {author}
  {\bibfnamefont {D.~V.}\ \bibnamefont {Averin}},\ }\href {\doibase
  10.1103/RevModPhys.85.1421} {\bibfield  {journal} {\bibinfo  {journal} {Rev.
  Mod. Phys.}\ }\textbf {\bibinfo {volume} {85}},\ \bibinfo {pages} {1421}
  (\bibinfo {year} {2013})}\BibitemShut {NoStop}%
\bibitem [{\citenamefont {Martinez-Perez}\ \emph {et~al.}(2015)\citenamefont
  {Martinez-Perez}, \citenamefont {Fornieri},\ and\ \citenamefont
  {Giazotto}}]{Giazotto}%
  \BibitemOpen
  \bibfield  {author} {\bibinfo {author} {\bibfnamefont {M.}~\bibnamefont
  {Martinez-Perez}}, \bibinfo {author} {\bibfnamefont {A.}~\bibnamefont
  {Fornieri}}, \ and\ \bibinfo {author} {\bibfnamefont {F.}~\bibnamefont
  {Giazotto}},\ }\href@noop {} {\bibfield  {journal} {\bibinfo  {journal}
  {Nature Nanotech.}\ }\textbf {\bibinfo {volume} {10}},\ \bibinfo {pages}
  {303} (\bibinfo {year} {2015})}\BibitemShut {NoStop}%
\bibitem [{\citenamefont {Nahum}\ \emph {et~al.}(1994)\citenamefont {Nahum},
  \citenamefont {Eiles},\ and\ \citenamefont {Martinis}}]{martiniscool}%
  \BibitemOpen
  \bibfield  {author} {\bibinfo {author} {\bibfnamefont {M.}~\bibnamefont
  {Nahum}}, \bibinfo {author} {\bibfnamefont {T.~M.}\ \bibnamefont {Eiles}}, \
  and\ \bibinfo {author} {\bibfnamefont {J.~M.}\ \bibnamefont {Martinis}},\
  }\href {\doibase 10.1063/1.112456} {\bibfield  {journal} {\bibinfo  {journal}
  {Applied Physics Letters}\ }\textbf {\bibinfo {volume} {65}},\ \bibinfo
  {pages} {3123} (\bibinfo {year} {1994})}\BibitemShut {NoStop}%
\bibitem [{\citenamefont {Muhonen}\ \emph {et~al.}(2012)\citenamefont
  {Muhonen}, \citenamefont {Meschke},\ and\ \citenamefont {Pekola}}]{muhonen}%
  \BibitemOpen
  \bibfield  {author} {\bibinfo {author} {\bibfnamefont {J.~T.}\ \bibnamefont
  {Muhonen}}, \bibinfo {author} {\bibfnamefont {M.}~\bibnamefont {Meschke}}, \
  and\ \bibinfo {author} {\bibfnamefont {J.~P.}\ \bibnamefont {Pekola}},\
  }\href {http://stacks.iop.org/0034-4885/75/i=4/a=046501} {\bibfield
  {journal} {\bibinfo  {journal} {Reports on Progress in Physics}\ }\textbf
  {\bibinfo {volume} {75}},\ \bibinfo {pages} {046501} (\bibinfo {year}
  {2012})}\BibitemShut {NoStop}%
\bibitem [{\citenamefont {Zhang}\ \emph {et~al.}(2015)\citenamefont {Zhang},
  \citenamefont {Lowell}, \citenamefont {Wilson}, \citenamefont {O'Neil},\ and\
  \citenamefont {Ullom}}]{ZhangUllom}%
  \BibitemOpen
  \bibfield  {author} {\bibinfo {author} {\bibfnamefont {X.}~\bibnamefont
  {Zhang}}, \bibinfo {author} {\bibfnamefont {P.~J.}\ \bibnamefont {Lowell}},
  \bibinfo {author} {\bibfnamefont {B.~L.}\ \bibnamefont {Wilson}}, \bibinfo
  {author} {\bibfnamefont {G.~C.}\ \bibnamefont {O'Neil}}, \ and\ \bibinfo
  {author} {\bibfnamefont {J.~N.}\ \bibnamefont {Ullom}},\ }\href {\doibase
  10.1103/PhysRevApplied.4.024006} {\bibfield  {journal} {\bibinfo  {journal}
  {Phys. Rev. Applied}\ }\textbf {\bibinfo {volume} {4}},\ \bibinfo {pages}
  {024006} (\bibinfo {year} {2015})}\BibitemShut {NoStop}%
\bibitem [{\citenamefont {Nguyen}\ \emph {et~al.}(2015)\citenamefont {Nguyen},
  \citenamefont {Meschke},\ and\ \citenamefont {Pekola}}]{Nguyen}%
  \BibitemOpen
  \bibfield  {author} {\bibinfo {author} {\bibfnamefont {H.~Q.}\ \bibnamefont
  {Nguyen}}, \bibinfo {author} {\bibfnamefont {M.}~\bibnamefont {Meschke}}, \
  and\ \bibinfo {author} {\bibfnamefont {J.~P.}\ \bibnamefont {Pekola}},\
  }\href {\doibase 10.1063/1.4905440} {\bibfield  {journal} {\bibinfo
  {journal} {Applied Physics Letters}\ }\textbf {\bibinfo {volume} {106}},\
  \bibinfo {pages} {012601} (\bibinfo {year} {2015})}\BibitemShut {NoStop}%
\bibitem [{\citenamefont {Mykk{\"a}nen}\ \emph {et~al.}(2020)\citenamefont
  {Mykk{\"a}nen}, \citenamefont {Lehtinen}, \citenamefont {Gr{\"o}nberg},
  \citenamefont {Shchepetov}, \citenamefont {Timofeev}, \citenamefont
  {Gunnarsson}, \citenamefont {Kemppinen}, \citenamefont {Manninen},\ and\
  \citenamefont {Prunnila}}]{Prunnila}%
  \BibitemOpen
  \bibfield  {author} {\bibinfo {author} {\bibfnamefont {E.}~\bibnamefont
  {Mykk{\"a}nen}}, \bibinfo {author} {\bibfnamefont {J.~S.}\ \bibnamefont
  {Lehtinen}}, \bibinfo {author} {\bibfnamefont {L.}~\bibnamefont
  {Gr{\"o}nberg}}, \bibinfo {author} {\bibfnamefont {A.}~\bibnamefont
  {Shchepetov}}, \bibinfo {author} {\bibfnamefont {A.~V.}\ \bibnamefont
  {Timofeev}}, \bibinfo {author} {\bibfnamefont {D.}~\bibnamefont
  {Gunnarsson}}, \bibinfo {author} {\bibfnamefont {A.}~\bibnamefont
  {Kemppinen}}, \bibinfo {author} {\bibfnamefont {A.~J.}\ \bibnamefont
  {Manninen}}, \ and\ \bibinfo {author} {\bibfnamefont {M.}~\bibnamefont
  {Prunnila}},\ }\href@noop {} {\ \textbf {\bibinfo {volume} {6}},\ \bibinfo
  {eid} {eaax9191} (\bibinfo {year} {2020})}\BibitemShut {NoStop}%
\bibitem [{\citenamefont {Koppinen}\ and\ \citenamefont
  {Maasilta}(2009)}]{PhysRevLett.102.165502}%
  \BibitemOpen
  \bibfield  {author} {\bibinfo {author} {\bibfnamefont {P.~J.}\ \bibnamefont
  {Koppinen}}\ and\ \bibinfo {author} {\bibfnamefont {I.~J.}\ \bibnamefont
  {Maasilta}},\ }\href {\doibase 10.1103/PhysRevLett.102.165502} {\bibfield
  {journal} {\bibinfo  {journal} {Phys. Rev. Lett.}\ }\textbf {\bibinfo
  {volume} {102}},\ \bibinfo {pages} {165502} (\bibinfo {year}
  {2009})}\BibitemShut {NoStop}%
\bibitem [{\citenamefont {Muhonen}\ \emph {et~al.}(2009)\citenamefont
  {Muhonen}, \citenamefont {Niskanen}, \citenamefont {Meschke}, \citenamefont
  {Pashkin}, \citenamefont {Tsai}, \citenamefont {Sainiemi}, \citenamefont
  {Franssila},\ and\ \citenamefont {Pekola}}]{Juha}%
  \BibitemOpen
  \bibfield  {author} {\bibinfo {author} {\bibfnamefont {J.~T.}\ \bibnamefont
  {Muhonen}}, \bibinfo {author} {\bibfnamefont {A.~O.}\ \bibnamefont
  {Niskanen}}, \bibinfo {author} {\bibfnamefont {M.}~\bibnamefont {Meschke}},
  \bibinfo {author} {\bibfnamefont {Y.~A.}\ \bibnamefont {Pashkin}}, \bibinfo
  {author} {\bibfnamefont {J.~S.}\ \bibnamefont {Tsai}}, \bibinfo {author}
  {\bibfnamefont {L.}~\bibnamefont {Sainiemi}}, \bibinfo {author}
  {\bibfnamefont {S.}~\bibnamefont {Franssila}}, \ and\ \bibinfo {author}
  {\bibfnamefont {J.~P.}\ \bibnamefont {Pekola}},\ }\href {\doibase
  10.1063/1.3080668} {\bibfield  {journal} {\bibinfo  {journal} {Applied
  Physics Letters}\ }\textbf {\bibinfo {volume} {94}},\ \bibinfo {pages}
  {073101} (\bibinfo {year} {2009})}\BibitemShut {NoStop}%
\bibitem [{\citenamefont {Miller}\ \emph {et~al.}(2008)\citenamefont {Miller},
  \citenamefont {O'Neil}, \citenamefont {Beall}, \citenamefont {Hilton},
  \citenamefont {Irwin}, \citenamefont {Schmidt}, \citenamefont {Vale},\ and\
  \citenamefont {Ullom}}]{Ullom}%
  \BibitemOpen
  \bibfield  {author} {\bibinfo {author} {\bibfnamefont {N.~A.}\ \bibnamefont
  {Miller}}, \bibinfo {author} {\bibfnamefont {G.~C.}\ \bibnamefont {O'Neil}},
  \bibinfo {author} {\bibfnamefont {J.~A.}\ \bibnamefont {Beall}}, \bibinfo
  {author} {\bibfnamefont {G.~C.}\ \bibnamefont {Hilton}}, \bibinfo {author}
  {\bibfnamefont {K.~D.}\ \bibnamefont {Irwin}}, \bibinfo {author}
  {\bibfnamefont {D.~R.}\ \bibnamefont {Schmidt}}, \bibinfo {author}
  {\bibfnamefont {L.~R.}\ \bibnamefont {Vale}}, \ and\ \bibinfo {author}
  {\bibfnamefont {J.~N.}\ \bibnamefont {Ullom}},\ }\href {\doibase
  10.1063/1.2913160} {\bibfield  {journal} {\bibinfo  {journal} {Applied
  Physics Letters}\ }\textbf {\bibinfo {volume} {92}},\ \bibinfo {pages}
  {163501} (\bibinfo {year} {2008})}\BibitemShut {NoStop}%
\bibitem [{\citenamefont {Timofeev}\ \emph {et~al.}(2009)\citenamefont
  {Timofeev}, \citenamefont {Helle}, \citenamefont {Meschke}, \citenamefont
  {M\"ott\"onen},\ and\ \citenamefont {Pekola}}]{PhysRevLett.102.200801}%
  \BibitemOpen
  \bibfield  {author} {\bibinfo {author} {\bibfnamefont {A.~V.}\ \bibnamefont
  {Timofeev}}, \bibinfo {author} {\bibfnamefont {M.}~\bibnamefont {Helle}},
  \bibinfo {author} {\bibfnamefont {M.}~\bibnamefont {Meschke}}, \bibinfo
  {author} {\bibfnamefont {M.}~\bibnamefont {M\"ott\"onen}}, \ and\ \bibinfo
  {author} {\bibfnamefont {J.~P.}\ \bibnamefont {Pekola}},\ }\href {\doibase
  10.1103/PhysRevLett.102.200801} {\bibfield  {journal} {\bibinfo  {journal}
  {Phys. Rev. Lett.}\ }\textbf {\bibinfo {volume} {102}},\ \bibinfo {pages}
  {200801} (\bibinfo {year} {2009})}\BibitemShut {NoStop}%
\bibitem [{\citenamefont {Partanen}\ \emph {et~al.}(2016)\citenamefont
  {Partanen}, \citenamefont {Tan}, \citenamefont {Govenius}, \citenamefont
  {Lake}, \citenamefont {M\"akel\"a}, \citenamefont {Tanttu},\ and\
  \citenamefont {M\"ott\"onen}}]{Mottonen}%
  \BibitemOpen
  \bibfield  {author} {\bibinfo {author} {\bibfnamefont {M.}~\bibnamefont
  {Partanen}}, \bibinfo {author} {\bibfnamefont {K.~Y.}\ \bibnamefont {Tan}},
  \bibinfo {author} {\bibfnamefont {J.}~\bibnamefont {Govenius}}, \bibinfo
  {author} {\bibfnamefont {R.~E.}\ \bibnamefont {Lake}}, \bibinfo {author}
  {\bibfnamefont {M.~K.}\ \bibnamefont {M\"akel\"a}}, \bibinfo {author}
  {\bibfnamefont {T.}~\bibnamefont {Tanttu}}, \ and\ \bibinfo {author}
  {\bibfnamefont {M.}~\bibnamefont {M\"ott\"onen}},\ }\href@noop {} {\bibfield
  {journal} {\bibinfo  {journal} {Nat. Phys.}\ }\textbf {\bibinfo {volume}
  {12}},\ \bibinfo {pages} {460} (\bibinfo {year} {2016})}\BibitemShut
  {NoStop}%
\bibitem [{\citenamefont {Tan}\ \emph {et~al.}(2017)\citenamefont {Tan},
  \citenamefont {Partanen}, \citenamefont {Lake}, \citenamefont {Govenius},
  \citenamefont {Masuda},\ and\ \citenamefont {M{\"o}tt{\"o}nen}}]{Qbit_sinis}%
  \BibitemOpen
  \bibfield  {author} {\bibinfo {author} {\bibfnamefont {K.~Y.}\ \bibnamefont
  {Tan}}, \bibinfo {author} {\bibfnamefont {M.}~\bibnamefont {Partanen}},
  \bibinfo {author} {\bibfnamefont {R.~E.}\ \bibnamefont {Lake}}, \bibinfo
  {author} {\bibfnamefont {J.}~\bibnamefont {Govenius}}, \bibinfo {author}
  {\bibfnamefont {S.}~\bibnamefont {Masuda}}, \ and\ \bibinfo {author}
  {\bibfnamefont {M.}~\bibnamefont {M{\"o}tt{\"o}nen}},\ }\href {\doibase
  10.1038/ncomms15189} {\bibfield  {journal} {\bibinfo  {journal} {Nature
  Communications}\ }\textbf {\bibinfo {volume} {8}},\ \bibinfo {pages} {15189}
  (\bibinfo {year} {2017})}\BibitemShut {NoStop}%
\bibitem [{\citenamefont {Sun}\ \emph {et~al.}(1999)\citenamefont {Sun},
  \citenamefont {Matsuo},\ and\ \citenamefont {Misawa}}]{sun}%
  \BibitemOpen
  \bibfield  {author} {\bibinfo {author} {\bibfnamefont {H.-B.}\ \bibnamefont
  {Sun}}, \bibinfo {author} {\bibfnamefont {S.}~\bibnamefont {Matsuo}}, \ and\
  \bibinfo {author} {\bibfnamefont {H.}~\bibnamefont {Misawa}},\ }\href
  {\doibase 10.1063/1.123367} {\bibfield  {journal} {\bibinfo  {journal}
  {Applied Physics Letters}\ }\textbf {\bibinfo {volume} {74}},\ \bibinfo
  {pages} {786} (\bibinfo {year} {1999})}\BibitemShut {NoStop}%
\bibitem [{\citenamefont {Kawata}\ \emph {et~al.}(2001)\citenamefont {Kawata},
  \citenamefont {Sun}, \citenamefont {Tanaka},\ and\ \citenamefont
  {Takada}}]{kawata}%
  \BibitemOpen
  \bibfield  {author} {\bibinfo {author} {\bibfnamefont {S.}~\bibnamefont
  {Kawata}}, \bibinfo {author} {\bibfnamefont {H.-B.}\ \bibnamefont {Sun}},
  \bibinfo {author} {\bibfnamefont {T.}~\bibnamefont {Tanaka}}, \ and\ \bibinfo
  {author} {\bibfnamefont {K.}~\bibnamefont {Takada}},\ }\href@noop {}
  {\bibfield  {journal} {\bibinfo  {journal} {Nature (London)}\ }\textbf
  {\bibinfo {volume} {412}},\ \bibinfo {pages} {697} (\bibinfo {year}
  {2001})}\BibitemShut {NoStop}%
\bibitem [{\citenamefont {Deubel}\ \emph {et~al.}(2004)\citenamefont {Deubel},
  \citenamefont {von Freymann}, \citenamefont {Wegener}, \citenamefont
  {Pereira}, \citenamefont {Busch},\ and\ \citenamefont {Soukoulis}}]{deubel}%
  \BibitemOpen
  \bibfield  {author} {\bibinfo {author} {\bibfnamefont {M.}~\bibnamefont
  {Deubel}}, \bibinfo {author} {\bibfnamefont {G.}~\bibnamefont {von
  Freymann}}, \bibinfo {author} {\bibfnamefont {M.}~\bibnamefont {Wegener}},
  \bibinfo {author} {\bibfnamefont {S.}~\bibnamefont {Pereira}}, \bibinfo
  {author} {\bibfnamefont {K.}~\bibnamefont {Busch}}, \ and\ \bibinfo {author}
  {\bibfnamefont {C.~M.}\ \bibnamefont {Soukoulis}},\ }\href@noop {} {\bibfield
   {journal} {\bibinfo  {journal} {Nature Materials}\ }\textbf {\bibinfo
  {volume} {3}},\ \bibinfo {pages} {444} (\bibinfo {year} {2004})}\BibitemShut
  {NoStop}%
\bibitem [{\citenamefont {Braun}\ and\ \citenamefont
  {Maier}(2016)}]{acssensors.6b00469}%
  \BibitemOpen
  \bibfield  {author} {\bibinfo {author} {\bibfnamefont {A.}~\bibnamefont
  {Braun}}\ and\ \bibinfo {author} {\bibfnamefont {S.~A.}\ \bibnamefont
  {Maier}},\ }\href {\doibase 10.1021/acssensors.6b00469} {\bibfield  {journal}
  {\bibinfo  {journal} {ACS Sensors}\ }\textbf {\bibinfo {volume} {1}},\
  \bibinfo {pages} {1155} (\bibinfo {year} {2016})}\BibitemShut {NoStop}%
\bibitem [{\citenamefont {Heiskanen}\ \emph {et~al.}(2020)\citenamefont
  {Heiskanen}, \citenamefont {Geng}, \citenamefont {Mastom\"aki},\ and\
  \citenamefont {Maasilta}}]{samuli}%
  \BibitemOpen
  \bibfield  {author} {\bibinfo {author} {\bibfnamefont {S.}~\bibnamefont
  {Heiskanen}}, \bibinfo {author} {\bibfnamefont {Z.}~\bibnamefont {Geng}},
  \bibinfo {author} {\bibfnamefont {J.}~\bibnamefont {Mastom\"aki}}, \ and\
  \bibinfo {author} {\bibfnamefont {I.~J.}\ \bibnamefont {Maasilta}},\ }\href
  {\doibase 10.1002/adem.201901290} {\bibfield  {journal} {\bibinfo  {journal}
  {Advanced Engineering Materials}\ }\textbf {\bibinfo {volume} {22}},\
  \bibinfo {pages} {1901290} (\bibinfo {year} {2020})}\BibitemShut {NoStop}%
\bibitem [{\citenamefont {Cooper}\ \emph {et~al.}(2007)\citenamefont {Cooper},
  \citenamefont {Hamel}, \citenamefont {Whitney}, \citenamefont {Weilermann},
  \citenamefont {Kramer}, \citenamefont {Zhao},\ and\ \citenamefont
  {Gentile}}]{spray}%
  \BibitemOpen
  \bibfield  {author} {\bibinfo {author} {\bibfnamefont {K.~A.}\ \bibnamefont
  {Cooper}}, \bibinfo {author} {\bibfnamefont {C.}~\bibnamefont {Hamel}},
  \bibinfo {author} {\bibfnamefont {B.}~\bibnamefont {Whitney}}, \bibinfo
  {author} {\bibfnamefont {K.}~\bibnamefont {Weilermann}}, \bibinfo {author}
  {\bibfnamefont {K.~J.}\ \bibnamefont {Kramer}}, \bibinfo {author}
  {\bibfnamefont {Y.}~\bibnamefont {Zhao}}, \ and\ \bibinfo {author}
  {\bibfnamefont {H.}~\bibnamefont {Gentile}},\ }\href@noop {} {\emph {\bibinfo
  {title} {Conformal photoresist coatings for high aspect ratio features}}}\
  (\bibinfo  {publisher} {SUSS MicroTec Waterbury Center},\ \bibinfo {address}
  {VT, USA},\ \bibinfo {year} {2007})\BibitemShut {NoStop}%
\bibitem [{\citenamefont {Dynes}\ \emph {et~al.}(1984)\citenamefont {Dynes},
  \citenamefont {Garno}, \citenamefont {Hertel},\ and\ \citenamefont
  {Orlando}}]{dynes}%
  \BibitemOpen
  \bibfield  {author} {\bibinfo {author} {\bibfnamefont {R.~C.}\ \bibnamefont
  {Dynes}}, \bibinfo {author} {\bibfnamefont {J.~P.}\ \bibnamefont {Garno}},
  \bibinfo {author} {\bibfnamefont {G.~B.}\ \bibnamefont {Hertel}}, \ and\
  \bibinfo {author} {\bibfnamefont {T.~P.}\ \bibnamefont {Orlando}},\ }\href
  {\doibase 10.1103/PhysRevLett.53.2437} {\bibfield  {journal} {\bibinfo
  {journal} {Phys. Rev. Lett.}\ }\textbf {\bibinfo {volume} {53}},\ \bibinfo
  {pages} {2437} (\bibinfo {year} {1984})}\BibitemShut {NoStop}%
\bibitem [{\citenamefont {Pekola}\ \emph {et~al.}(2010)\citenamefont {Pekola},
  \citenamefont {Maisi}, \citenamefont {Kafanov}, \citenamefont {Chekurov},
  \citenamefont {Kemppinen}, \citenamefont {Pashkin}, \citenamefont {Saira},
  \citenamefont {M\"ott\"onen},\ and\ \citenamefont {Tsai}}]{Pekola_dynes}%
  \BibitemOpen
  \bibfield  {author} {\bibinfo {author} {\bibfnamefont {J.~P.}\ \bibnamefont
  {Pekola}}, \bibinfo {author} {\bibfnamefont {V.~F.}\ \bibnamefont {Maisi}},
  \bibinfo {author} {\bibfnamefont {S.}~\bibnamefont {Kafanov}}, \bibinfo
  {author} {\bibfnamefont {N.}~\bibnamefont {Chekurov}}, \bibinfo {author}
  {\bibfnamefont {A.}~\bibnamefont {Kemppinen}}, \bibinfo {author}
  {\bibfnamefont {Y.~A.}\ \bibnamefont {Pashkin}}, \bibinfo {author}
  {\bibfnamefont {O.-P.}\ \bibnamefont {Saira}}, \bibinfo {author}
  {\bibfnamefont {M.}~\bibnamefont {M\"ott\"onen}}, \ and\ \bibinfo {author}
  {\bibfnamefont {J.~S.}\ \bibnamefont {Tsai}},\ }\href {\doibase
  10.1103/PhysRevLett.105.026803} {\bibfield  {journal} {\bibinfo  {journal}
  {Phys. Rev. Lett.}\ }\textbf {\bibinfo {volume} {105}},\ \bibinfo {pages}
  {026803} (\bibinfo {year} {2010})}\BibitemShut {NoStop}%
\bibitem [{\citenamefont {Wellstood}\ \emph {et~al.}(1994)\citenamefont
  {Wellstood}, \citenamefont {Urbina},\ and\ \citenamefont
  {Clarke}}]{Wellstood}%
  \BibitemOpen
  \bibfield  {author} {\bibinfo {author} {\bibfnamefont {F.~C.}\ \bibnamefont
  {Wellstood}}, \bibinfo {author} {\bibfnamefont {C.}~\bibnamefont {Urbina}}, \
  and\ \bibinfo {author} {\bibfnamefont {J.}~\bibnamefont {Clarke}},\ }\href
  {\doibase 10.1103/PhysRevB.49.5942} {\bibfield  {journal} {\bibinfo
  {journal} {Phys. Rev. B}\ }\textbf {\bibinfo {volume} {49}},\ \bibinfo
  {pages} {5942} (\bibinfo {year} {1994})}\BibitemShut {NoStop}%
\bibitem [{\citenamefont {Pekola}\ \emph {et~al.}(2004)\citenamefont {Pekola},
  \citenamefont {Heikkil\"a}, \citenamefont {Savin}, \citenamefont {Flyktman},
  \citenamefont {Giazotto},\ and\ \citenamefont
  {Hekking}}]{PhysRevLett.92.056804}%
  \BibitemOpen
  \bibfield  {author} {\bibinfo {author} {\bibfnamefont {J.~P.}\ \bibnamefont
  {Pekola}}, \bibinfo {author} {\bibfnamefont {T.~T.}\ \bibnamefont
  {Heikkil\"a}}, \bibinfo {author} {\bibfnamefont {A.~M.}\ \bibnamefont
  {Savin}}, \bibinfo {author} {\bibfnamefont {J.~T.}\ \bibnamefont {Flyktman}},
  \bibinfo {author} {\bibfnamefont {F.}~\bibnamefont {Giazotto}}, \ and\
  \bibinfo {author} {\bibfnamefont {F.~W.~J.}\ \bibnamefont {Hekking}},\ }\href
  {\doibase 10.1103/PhysRevLett.92.056804} {\bibfield  {journal} {\bibinfo
  {journal} {Phys. Rev. Lett.}\ }\textbf {\bibinfo {volume} {92}},\ \bibinfo
  {pages} {056804} (\bibinfo {year} {2004})}\BibitemShut {NoStop}%
\bibitem [{\citenamefont {Pekola}\ \emph {et~al.}(2000)\citenamefont {Pekola},
  \citenamefont {Anghel}, \citenamefont {Suppula}, \citenamefont {Suoknuuti},
  \citenamefont {Manninen},\ and\ \citenamefont {Manninen}}]{traps}%
  \BibitemOpen
  \bibfield  {author} {\bibinfo {author} {\bibfnamefont {J.~P.}\ \bibnamefont
  {Pekola}}, \bibinfo {author} {\bibfnamefont {D.~V.}\ \bibnamefont {Anghel}},
  \bibinfo {author} {\bibfnamefont {T.~I.}\ \bibnamefont {Suppula}}, \bibinfo
  {author} {\bibfnamefont {J.~K.}\ \bibnamefont {Suoknuuti}}, \bibinfo {author}
  {\bibfnamefont {A.~J.}\ \bibnamefont {Manninen}}, \ and\ \bibinfo {author}
  {\bibfnamefont {M.}~\bibnamefont {Manninen}},\ }\href {\doibase
  10.1063/1.126474} {\bibfield  {journal} {\bibinfo  {journal} {Applied Physics
  Letters}\ }\textbf {\bibinfo {volume} {76}},\ \bibinfo {pages} {2782}
  (\bibinfo {year} {2000})}\BibitemShut {NoStop}%
\bibitem [{\citenamefont {Fisher}\ \emph {et~al.}(1999)\citenamefont {Fisher},
  \citenamefont {Ullom},\ and\ \citenamefont {Nahum}}]{fisherullom}%
  \BibitemOpen
  \bibfield  {author} {\bibinfo {author} {\bibfnamefont {P.~A.}\ \bibnamefont
  {Fisher}}, \bibinfo {author} {\bibfnamefont {J.~N.}\ \bibnamefont {Ullom}}, \
  and\ \bibinfo {author} {\bibfnamefont {M.}~\bibnamefont {Nahum}},\ }\href
  {\doibase 10.1063/1.123943} {\bibfield  {journal} {\bibinfo  {journal}
  {Applied Physics Letters}\ }\textbf {\bibinfo {volume} {74}},\ \bibinfo
  {pages} {2705} (\bibinfo {year} {1999})}\BibitemShut {NoStop}%
\bibitem [{\citenamefont {Matthias}\ \emph {et~al.}(1963)\citenamefont
  {Matthias}, \citenamefont {Geballe},\ and\ \citenamefont
  {Compton}}]{RevModPhys.35.1}%
  \BibitemOpen
  \bibfield  {author} {\bibinfo {author} {\bibfnamefont {B.~T.}\ \bibnamefont
  {Matthias}}, \bibinfo {author} {\bibfnamefont {T.~H.}\ \bibnamefont
  {Geballe}}, \ and\ \bibinfo {author} {\bibfnamefont {V.~B.}\ \bibnamefont
  {Compton}},\ }\href {\doibase 10.1103/RevModPhys.35.1} {\bibfield  {journal}
  {\bibinfo  {journal} {Rev. Mod. Phys.}\ }\textbf {\bibinfo {volume} {35}},\
  \bibinfo {pages} {1} (\bibinfo {year} {1963})}\BibitemShut {NoStop}%
\end{thebibliography}
%\bibliographystyle{plain}

\end{document}